\documentclass[aip,jap,reprint]{revtex4-2}
\usepackage{amsmath,amssymb}
\usepackage[utf8]{inputenc}
\usepackage{graphicx}
\usepackage{xcolor}
\usepackage{ulem}
\usepackage{dcolumn}
\usepackage{hyperref}
\usepackage{multirow}
\usepackage{placeins}
\usepackage{natbib}

\begin{document}

\preprint{AIP/123-QED}
\title[Neural-Network K-shell ionization]{IKEBANA: A Neural-Network approach for the K-shell ionization by electron impact}

\author{D.M. Mitnik} 
\email{dmitnik@df.uba.ar}
\author{C. C. Montanari} 
\affiliation{Instituto de Astronom\'{\i}a y F\'{\i}sica del Espacio, 
CONICET and Universidad de Buenos Aires, Buenos Aires, Argentina.}

\author{S.  Segui} 
\author{S. P. Limandri} 
\author{J. A. Guzm\'an} 
\author{A. C. Carreras} 
\author{J. C. Trincavelli} 
\affiliation{Instituto de Física Enrique Gaviola, CONICET and Facultad de Matemática, Astronomía, Física y Computación, Universidad Nacional de Córdoba, Córdoba, Argentina.}  
\makeatletter
\def\@email#1#2{%
 \endgroup
 \patchcmd{\titleblock@produce}
  {\frontmatter@RRAPformat}
  {\frontmatter@RRAPformat{\produce@RRAP{*#1\href{mailto:#2}{#2}}}\frontmatter@RRAPformat}
  {}{}
}%
\makeatother

\date{\today}


\begin{abstract}
A fully connected neural network was trained to model the K-shell ionization cross sections based on two input features: the atomic number and the incoming electron overvoltage.  
The training utilized a recent, updated compilation of experimental data \cite{database:24}, covering elements from H to U, 
and incident electron energies ranging from the threshold to relativistic values.
The neural network demonstrated excellent predictive performance, compared with the experimental data, when available, and with full theoretical predictions. 
The developed model is provided in the {\sc ikebana} code, which is openly available and requires only the user-selected atomic number and electron energy range as inputs. 
\end{abstract}



\maketitle
\section{Introduction}
\label{sec:introduction}

Electron-impact (EI) ionization has been a central topic in atomic and molecular physics since the advent of quantum mechanics. Its relevance spans a wide range of applications, including plasma diagnostics, astrophysical modeling, material characterization, mass spectrometry, and the fragmentation of complex biological molecules in radiotherapy and other medical treatments. Among the various parameters governing electron–matter interactions, the K-shell ionization cross section, $\sigma_K$, plays a pivotal role. Accurate knowledge of $\sigma_K$ is essential for interpreting X-ray emission spectra, understanding energy deposition processes in matter, and developing predictive models in both fundamental and applied physics. Despite nearly a century of theoretical and experimental efforts, precise and comprehensive data for these relevant parameters remain unknown. 

A recent and updated 2024 compilation of experimental $\sigma_K$ data~\cite{database:24}, covering elements from hydrogen to uranium, reveals that the majority of available measurements are concentrated in just eight elements: H, He, Ar, Cr, Fe, Ni, Cu, and Ag. In contrast, most other elements have only sparse data, typically limited to narrow energy intervals, and for approximately 30\% of the elements within the considered atomic number range, no experimental data are available at all.
This comprehensive dataset highlights both the scarcity of measurements for many elements and the significant variability among existing data, underscoring the need for reliable predictive models across the entire periodic table. 

Finding a universal prediction to accurately describe the EI ionization cross-sections of distinct target materials and along an extended energy range is a formidable challenge.
In this work, we developed a machine learning (ML) method based on a deep neural network, which can fulfill this task. 
When dealing with large volumes of data or seeking trends and patterns in complex datasets, machine learning approaches become a valuable tool~\cite{hao22,jaderberg24}. In particular, ML has proven to be useful in atomic physics research; for instance, a neural network trained with a complete set of experimental results has been applied to predict accurate electronic stopping power cross sections for any ion and target combination in a wide range of incident energies~\cite{haiek22}.

 The work is organized as follows: in Section \ref{sec:model} we describe the neural network model employed; in Section \ref{sec:results} the present results are displayed and discussed; and in Section \ref{sec:conclusions} we summarize the findings of the model and the conclusions arrived at. 

\section{Neural Network models}
\label{sec:model} 
A fully connected neural network was implemented to model the K-shell ionization cross-sections,  $\sigma_K$, based on two input features: the atomic number $Z$ and the incoming electron overvoltage
\begin{equation}
U_i \equiv \frac{E_i}{I_k} \, ,
\end{equation}
where $E_i$ is the incoming electron energy and $I_k$ is the K-shell ionization energy. 
The global structure of the neural network and the general training procedure applied follow the lineaments detailed in reference~\cite{haiek22}, although no data cleansing was implemented in the present work.
Since the involved values spanned wide ranges, the input features and the target variable were scaled before training. First, both $U$ and the ionization cross-sections $\sigma_K$ were transformed to a logarithmic scale (the atomic number $Z$ was kept in its original linear scale, as its logarithmic transformation did not improve the performance). Then, all the variables (input and output) were scaled using the \texttt{StandardScaler} function. The network architecture was composed of four layers: three hidden layers using \texttt{tanh} activation functions (as justified later), with $L2$ regularization ($\lambda = 10^{-5}$), batch normalization, and a small dropout rate ($10^{-4}$), followed by a final output layer with linear activation to regress the target variable. The architecture was as follows:
\begin{itemize}
    \item Input: 2 neurons ($Z$, $\log(U)$),
    \item Dense (32 units) $\rightarrow$ BatchNorm $\rightarrow$ Dropout,
    \item Dense (16 units) $\rightarrow$ BatchNorm $\rightarrow$ Dropout,
    \item Dense (32 units) $\rightarrow$ BatchNorm $\rightarrow$ Dropout,
    \item Output: 1 neuron (linear).
\end{itemize}

Our model separates 20\% of the available data as a test set, which is not used during training. From the remaining data, 20\% is allocated to the validation set. The network was trained using the adaptive moment estimation ({\sc adam}) optimizer with an initial learning rate of $3 \times 10^{-3}$, minimizing the mean squared error (MSE) and tracking the mean absolute error (MAE) as an auxiliary metric. Training was performed over 350 epochs with a batch size of 64, employing early stopping (patience = 60) and learning rate reduction on plateau (factor = 0.8, patience = 45), both based on the validation loss.

After evaluating several configurations, we selected the final architecture that defines the model used in our computational framework {\sc ikebana} ({\it Ionization of K-shell Electrons By A Neural-network Approach}).
Initially, \texttt{ReLU} activation functions were used, yielding highly accurate predictions. However, the resulting curves showed discontinuities and unphysical sharp features. Since electron-impact ionization cross-sections are expected to vary smoothly with energy, we first attempted to correct these artifacts by applying a Savitzky–Golay smoothing filter \cite{Savitzky:64} to the output. While this post-processing step successfully removed irregularities, it felt artificial, as smoothness was imposed externally rather than emerging from the model itself.
To address this in a more principled way, we tested a physics-informed variant of the network by introducing a penalty on the second derivative of the output with respect to $U$, aiming to encourage smoothness through the loss function. However, this approach did not improve performance and often resulted in worse outcomes. A two-phase strategy, involving initial training followed by fine-tuning with a smoothness penalty, was also tested but proved similarly ineffective.
Ultimately, replacing the activation functions with \texttt{tanh} resolved the issue. This change naturally led to continuous, physically realistic predictions without the need for external smoothing or additional loss terms.

\begin{figure}[h ]
\includegraphics[width=0.4\textwidth]{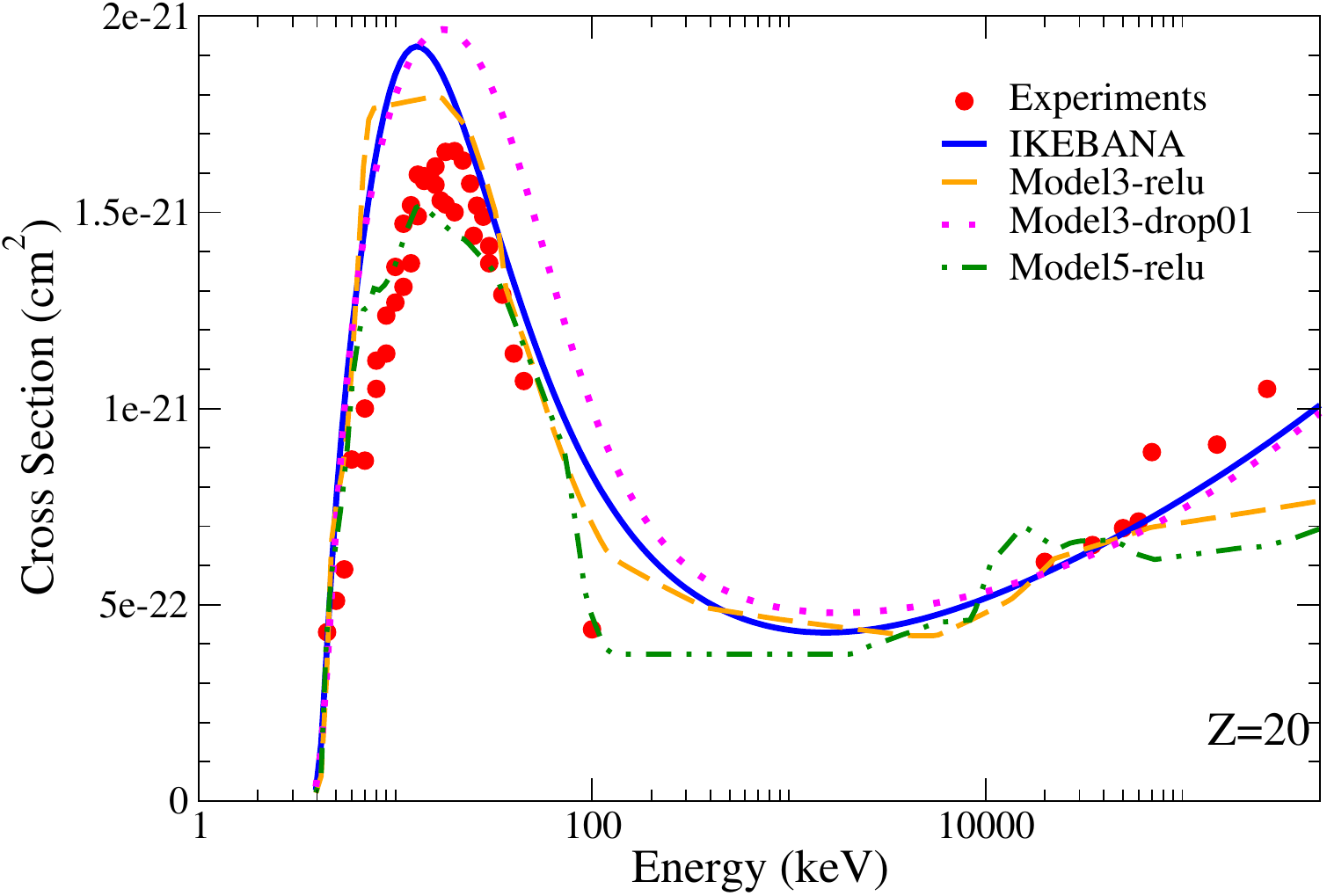}
\caption{\raggedright Predicted values of $\sigma_K$ for calcium using different neural network models, compared to the results from the final model {\sc ikebana}. The differences among the models are discussed in the text.}
\label{fig:model_comparison}
\end{figure}

While it is difficult to illustrate the motivations behind our architectural and training decisions with a single example, Fig.~\ref{fig:model_comparison} presents representative predictions obtained with different network variants for the case of $\sigma_K$ for calcium. Among the models tested, we selected three examples that enable direct comparison across architectures with differing activation functions, regularization parameters, and training strategies.
The {\sc model3-relu} differs from the final {\sc ikebana} model only in its use of \texttt{ReLU} activations. As shown, this model provides highly accurate predictions but produces jagged, non-smooth curves that fail to reproduce the physical behavior of the process. The second variant, {\sc model-3drop01}, shares the same architecture and hyperparameters as our final model, except for a significantly increased dropout rate (from $10^{-4}$ to $0.1$). This version improves predictions at high (relativistic) energies but overestimates cross-sections at low energies. While the position of the peak is closer to experimental values in this specific case, the overall deviation is larger.
The final example, {\sc model5-relu}, adds two more hidden layers and alters the architecture to 64-32-32-64-32-1, again using \texttt{ReLU} activations. This configuration yields overfitted predictions that consistently overshoot experimental values across the entire energy range.

These comparisons highlight how seemingly minor architectural decisions, such as the selection of activation functions or regularization strategies, can significantly impact the continuity and physical realism of the predictions. This comparative analysis played a key role in determining the optimal hyperparameter configuration for the final model.

\begin{figure}[h ]
\includegraphics[width=0.4\textwidth]{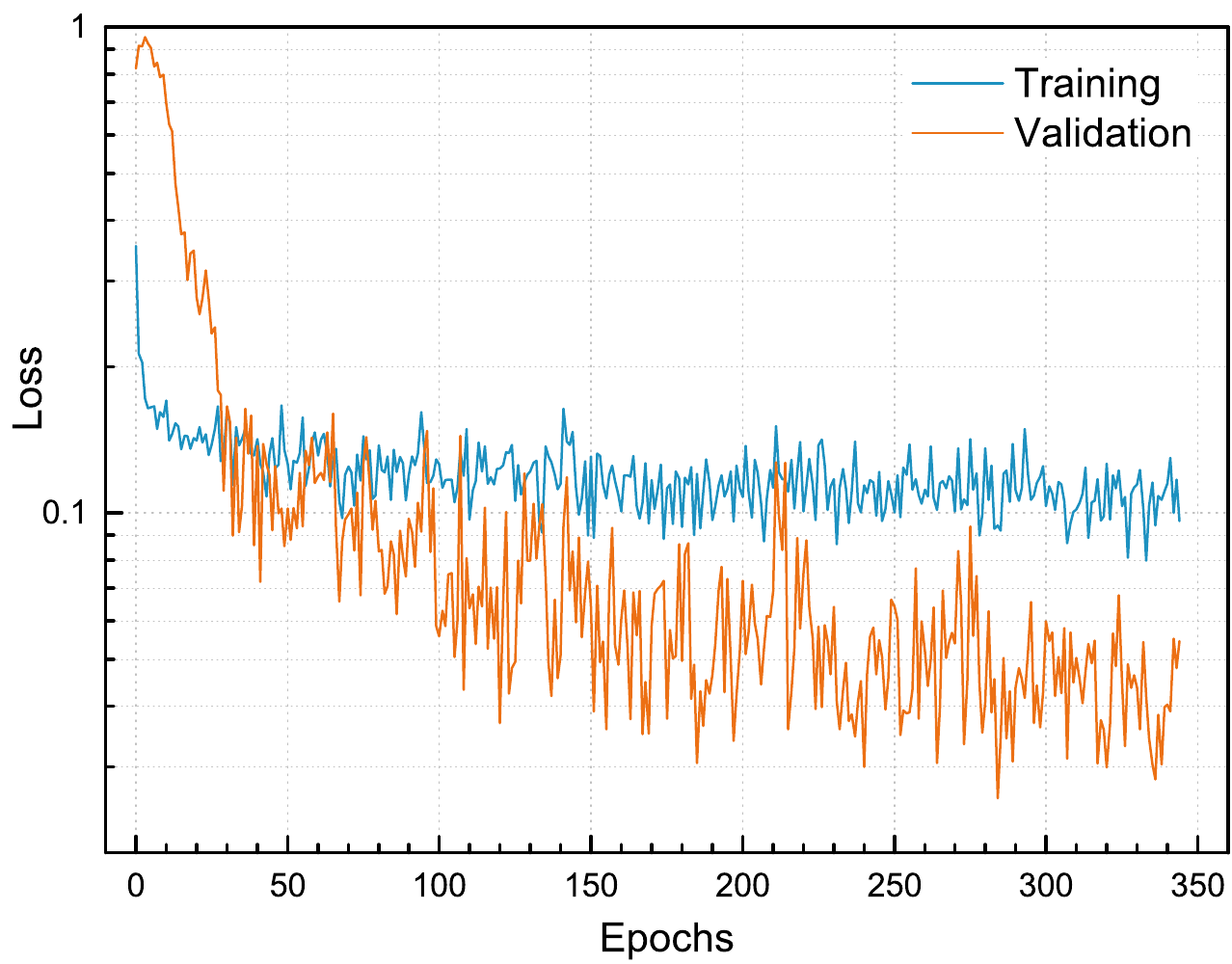}
\caption{\raggedright Evolution of the training and validation loss over epochs.}
\label{fig:training_history}
\end{figure}

The training and validation loss curves of our final {\sc ikebana} model  (Fig.~\ref{fig:training_history}) exhibit a consistent decrease and stabilization, with the validation loss consistently being lower than the training loss. This is likely a result of regularization techniques, such as dropout and batch normalization, which slightly underfit the training data while improving generalization. Early stopping also contributed to this behavior by preserving the model with the lowest validation loss.

\section{Results and discussion}
\label{sec:results}

The updated database used in this work \cite{database:24} contains over 2,500 experimental data points, compiled from approximately one hundred publications spanning from 1930 to 2024 (although no new experimental results were published after 2020). The reported cross sections cover a wide range of energies, although most measurements are concentrated between the ionization threshold and $U = 4$. A few experiments extend into the relativistic regime, with data reaching up to a few GeV. For some elements, only one or two data points are available.
A notable aspect of the dataset is the presence of significant discrepancies among results reported by different authors under the same conditions. At this stage, we chose not to filter out these inconsistencies, as was done in Ref.~\cite{haiek22}, to avoid making arbitrary decisions and to minimize the risk of overfitting during training.

\begin{figure*}[hbt] 
    \centering
    \includegraphics[width=0.3\textwidth]{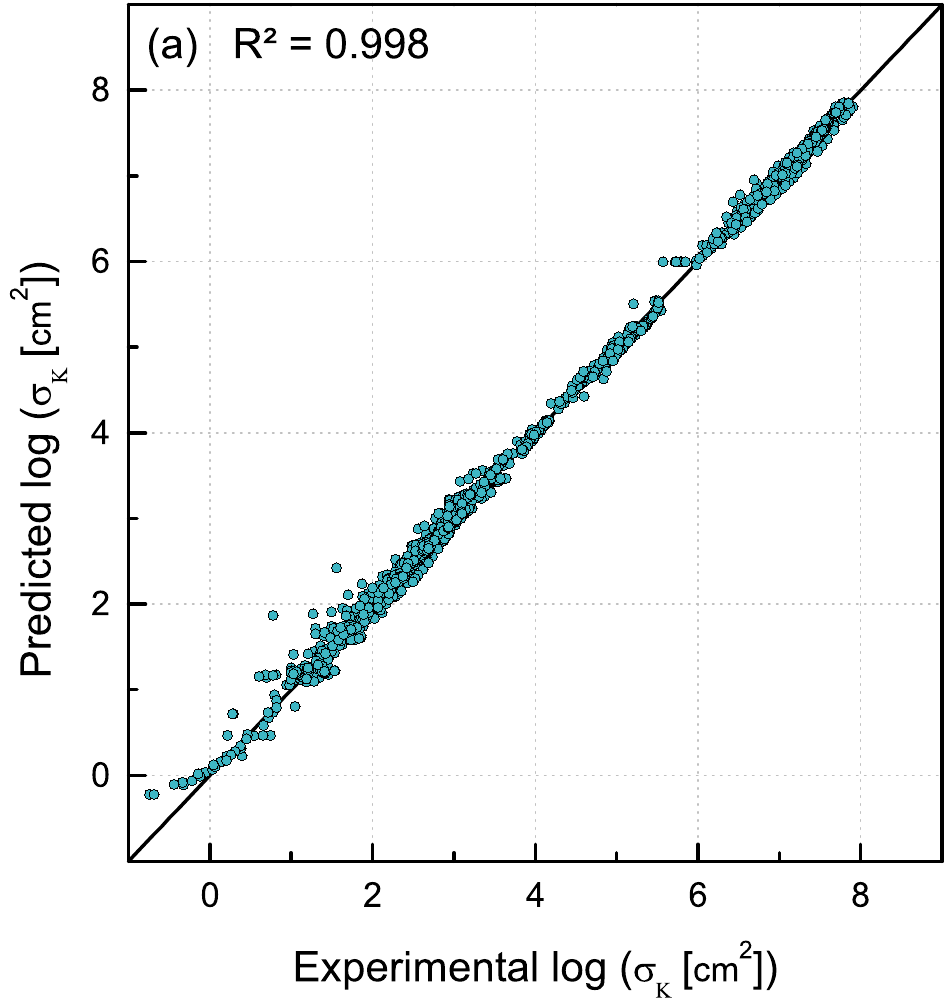}
    \hspace{0.03\textwidth}
    \includegraphics[width=0.3\textwidth]{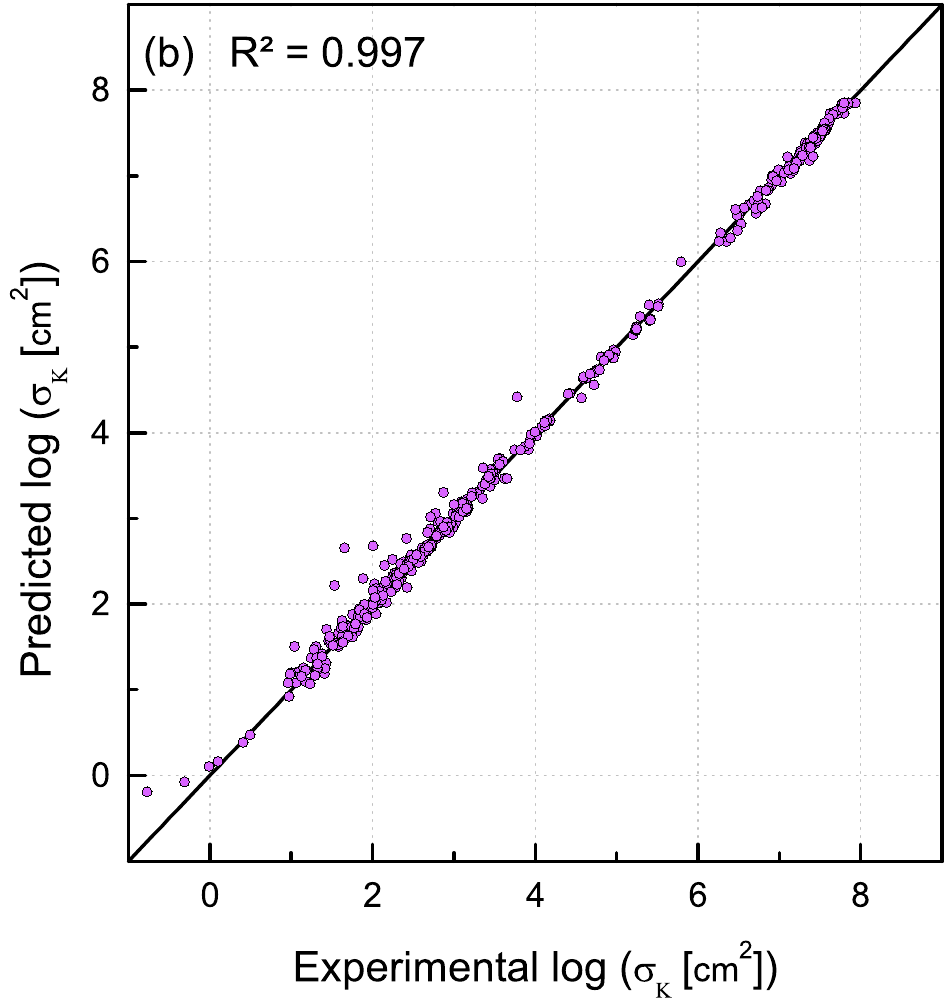}
    \caption{\raggedright Predicted vs experimental ionization cross-sections. (a) Training dataset; (b) test dataset.}
    \label{recta}
\end{figure*}

The present neural network code {\sc ikebana} demonstrates excellent predictive performance, and its results are compared in Fig.~\ref{recta} with the experimental values for the training and test datasets. The coefficient of determination reached $R^2 = 0.998$ in the training set and $R^2 = 0.997$ in the test set, indicating a strong generalization and minimal overfitting.

The goodness of the {\sc ikebana} model developed can also be assessed in terms of the relative deviation of the predicted values from the experimental ones. Different analyses can be performed in this regard, using either training or test datasets to study the values of $\sigma_K$ or $\log\sigma_K$. In addition, it is possible to take into account all the data or to exclude the values that deviate the most from the general trend. In Table~\ref{estadistica}, the relevant statistical parameters of this analysis are shown. 

\begin{table*}[hbt]
\caption{\label{estadistica} Statistical parameters of the distribution of residuals of {\sc ikebana} for $\sigma_K$ and $\log\sigma_K$.}
\centering
\begin{tabular}{|c|c||c|c||c|c|}
\hline
\multicolumn{2}{|c||}{} & \multicolumn{2}{c||}{Training set} & \multicolumn{2}{c|}{Test set} \\
\cline{3-6}
\multicolumn{2}{|c||}{} & full data & w/o outliers & full data & w/o outliers \\
\hline
         & Mean value               &   5   &   3   &   9   &   5   \\
$\sigma_K$ & Std. deviation (\%)      &  39   &  19   &  56   &  23   \\
         & Min. value               & -52   & -36   & -41   & -31   \\
         & Max. value               &1144   & 109   & 910   & 174   \\
\hline
         & Mean value               & -0.04 & -0.03 & -0.08 & -0.06 \\
$\log\sigma_K$ & Std. deviation (\%)  & 0.44  & 0.36  & 0.52  & 0.38  \\
         & Min. value               & -4.71 & -1.57 & -4.49 & -2.05 \\
         & Max. value               & 1.40  & 1.01  & 1.12  & 0.86  \\
\hline
\end{tabular}
\end{table*}

\begin{figure*}[h ] 
\includegraphics[width=0.42\textwidth]{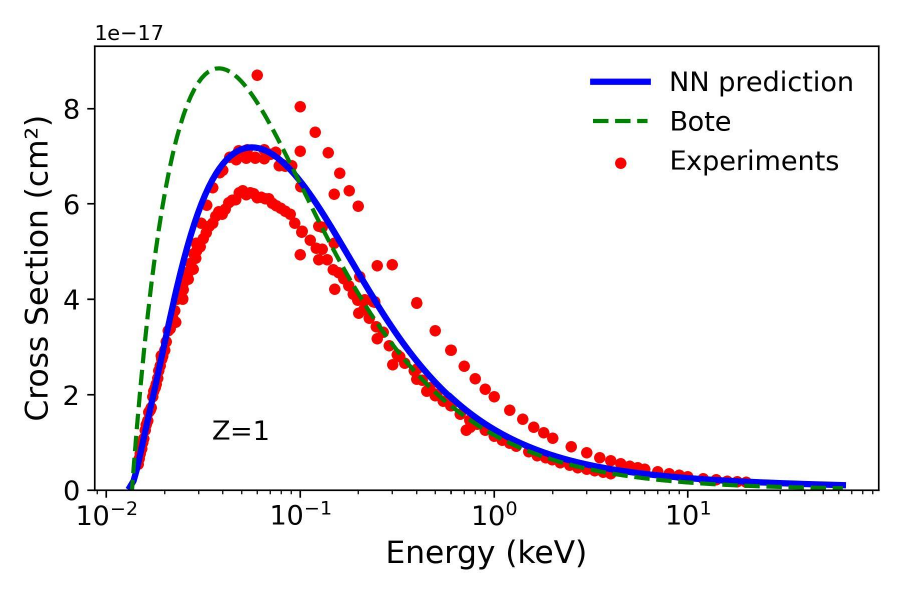}
\hspace{0.02\textwidth}
\includegraphics[width=0.42\textwidth]{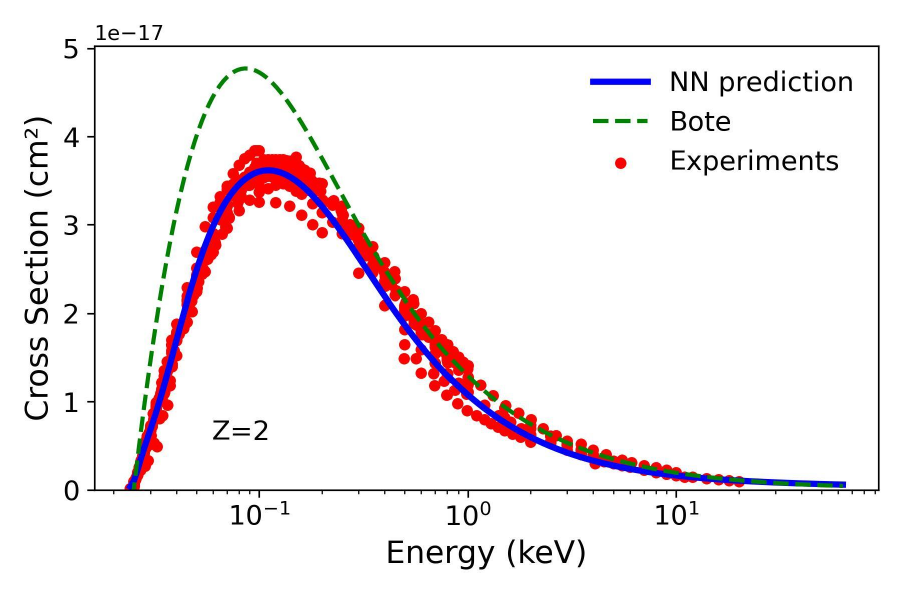} \\
\includegraphics[width=0.42\textwidth]{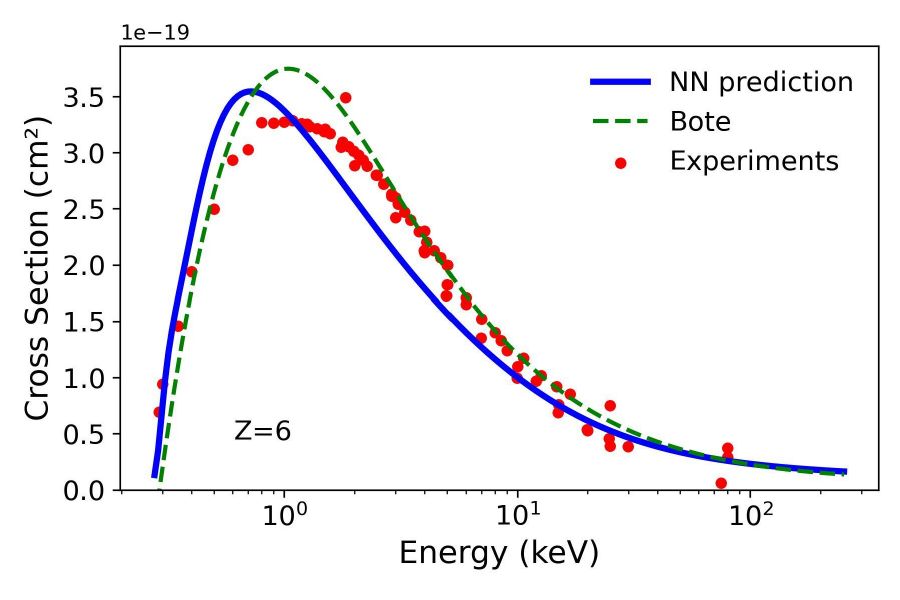}
\hspace{0.02\textwidth}
\includegraphics[width=0.42\textwidth]{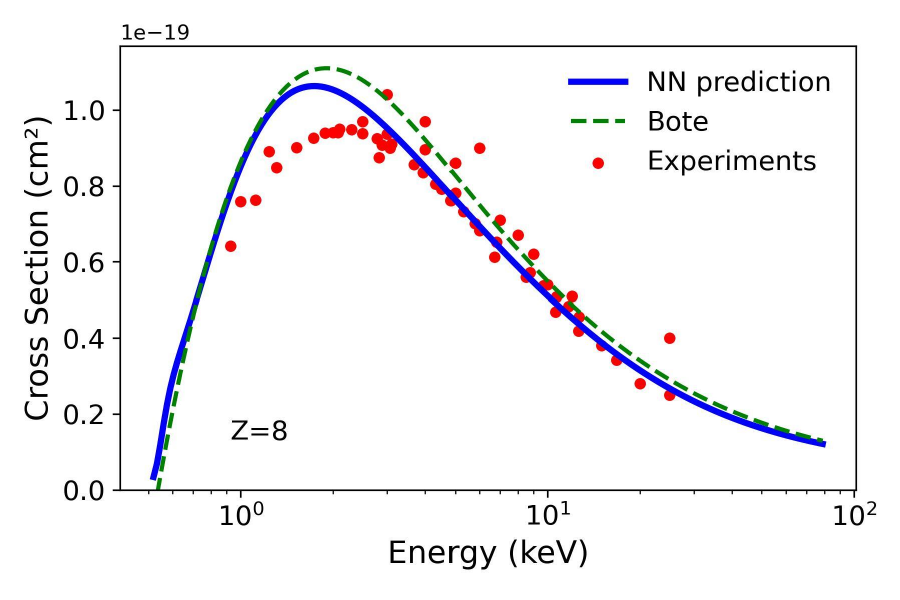} 
\caption{ \raggedright K-shell ionization cross sections as a function of the electron impact-energy for H, He, C, and O. 
A thick solid line is used for the values predicted by the neural-network code, whereas the dotted line represents the calculations performed by Bote {\it et al.}\cite{bote09}. The experimental results compiled in \cite{database:24} are shown as circles.}
\label{fig:KlowZ}
\end{figure*}

\begin{figure*}[h ] 
\includegraphics[width=0.42\textwidth]{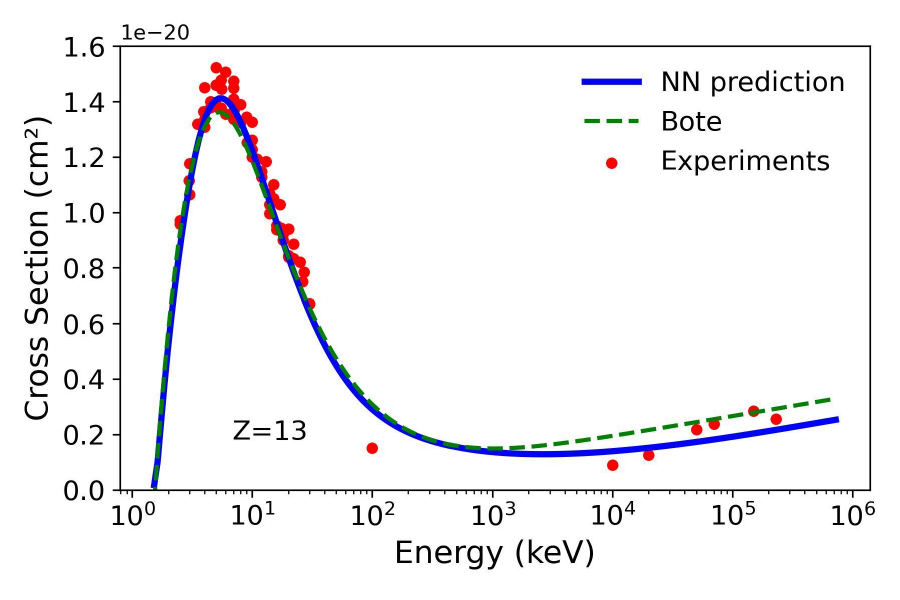}
\hspace{0.02\textwidth}
\includegraphics[width=0.42\textwidth]{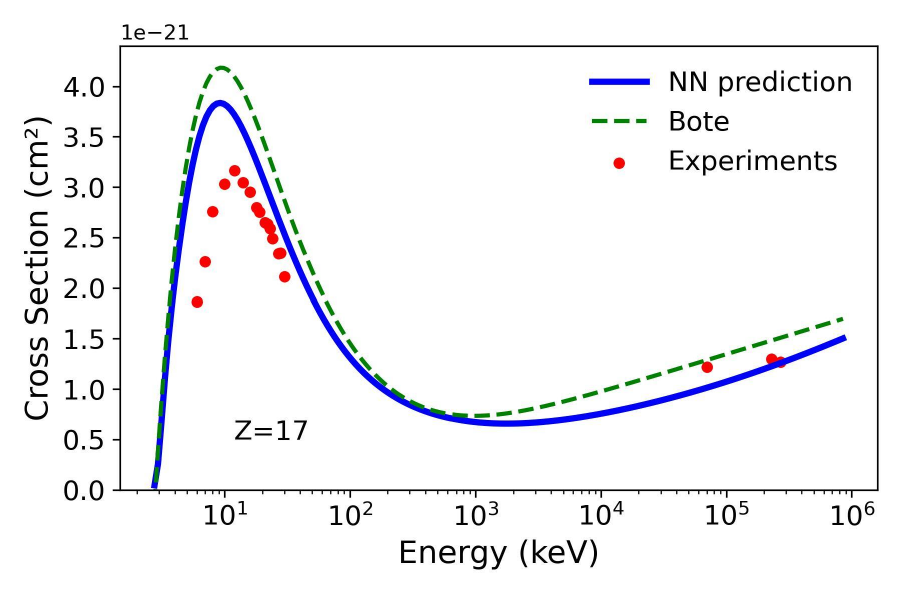} \\
\includegraphics[width=0.42\textwidth]{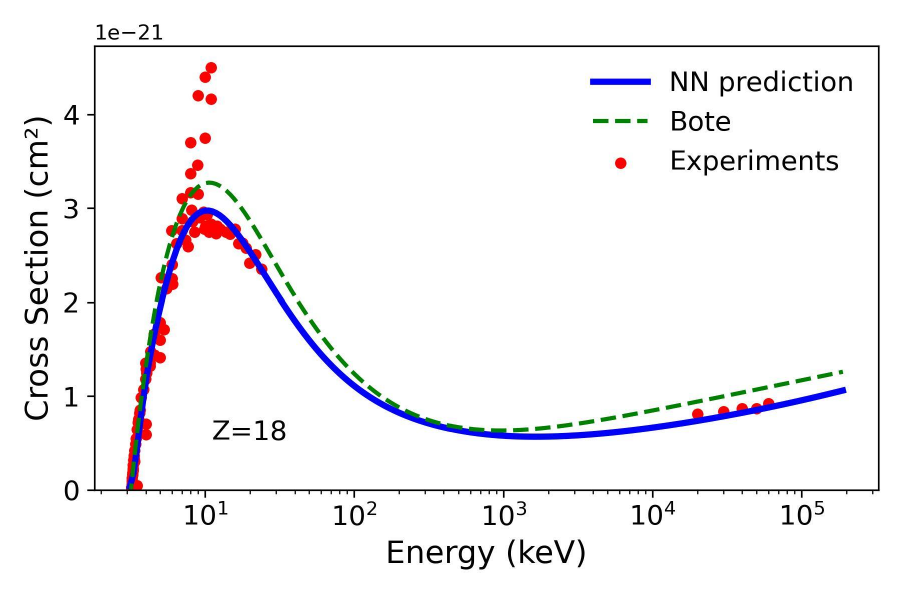}
\hspace{0.02\textwidth}
\includegraphics[width=0.42\textwidth]{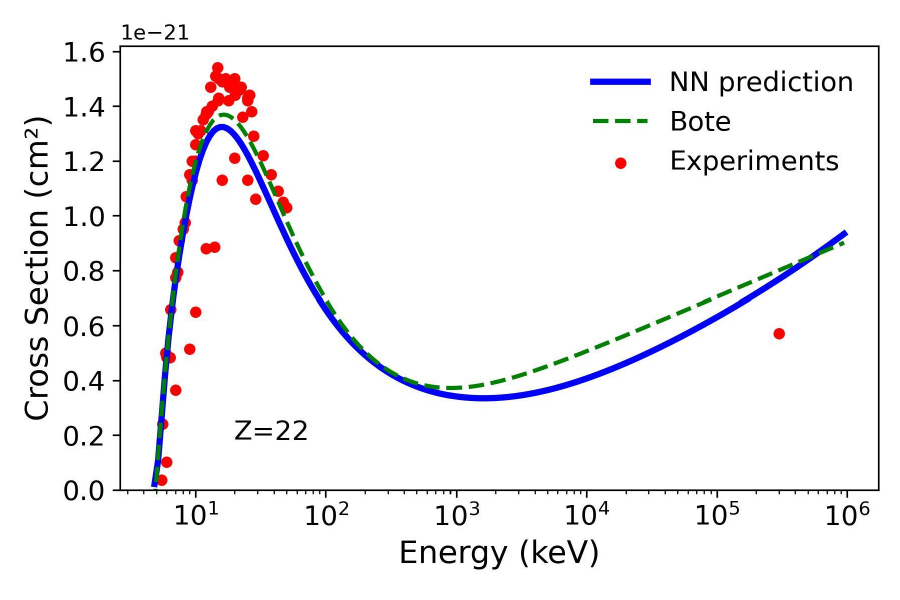} 
\caption{ \raggedright K-shell ionization cross sections as a function of the electron impact-energy for Al, Cl, Ar, and Ti. 
Curves and symbols as in Figure \ref{fig:KlowZ}.}
\label{fig:KAlClArCa}
\end{figure*}

\begin{figure*}[h ] 
\includegraphics[width=0.42\textwidth]{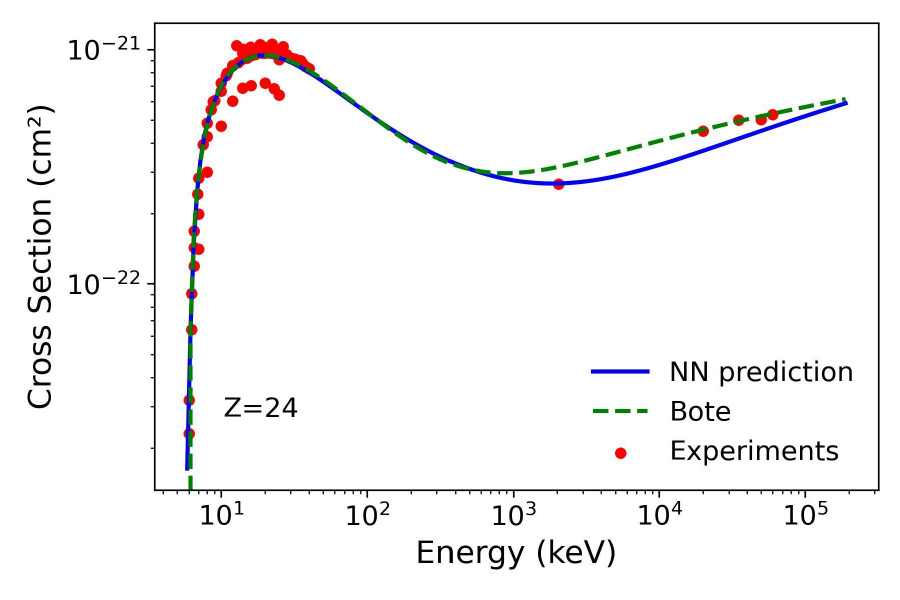}
\hspace{0.02\textwidth}
\includegraphics[width=0.42\textwidth]{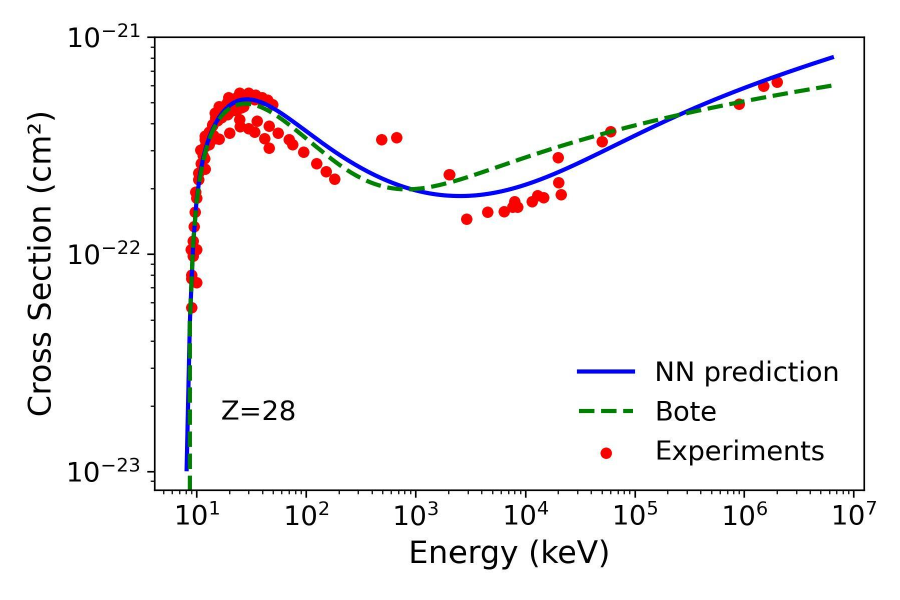} \\
\includegraphics[width=0.42\textwidth]{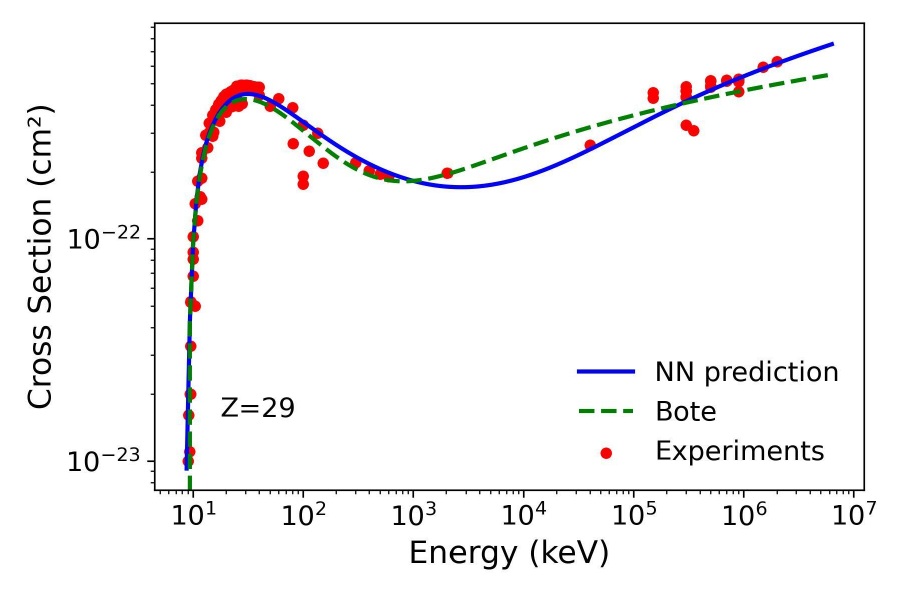}
\hspace{0.02\textwidth}
\includegraphics[width=0.42\textwidth]{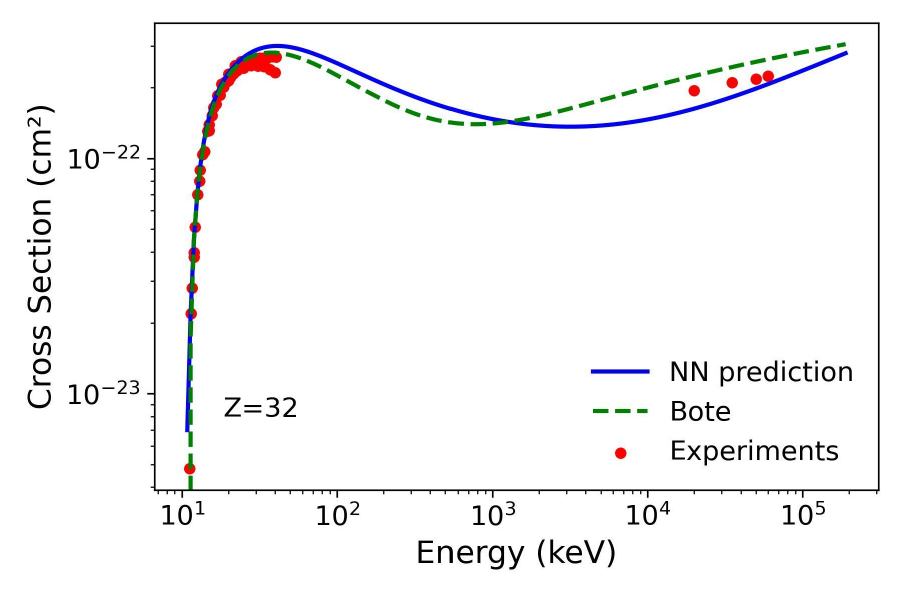} 
\caption{ \raggedright K-shell ionization cross sections as a function of the electron impact-energy for Cr, Ni, Cu, and Ge. 
Curves and symbols as in Figure \ref{fig:KlowZ}.}
\label{fig:KCrNiCuGe}
\end{figure*}

\begin{figure*}[h ] 
\includegraphics[width=0.42\textwidth]{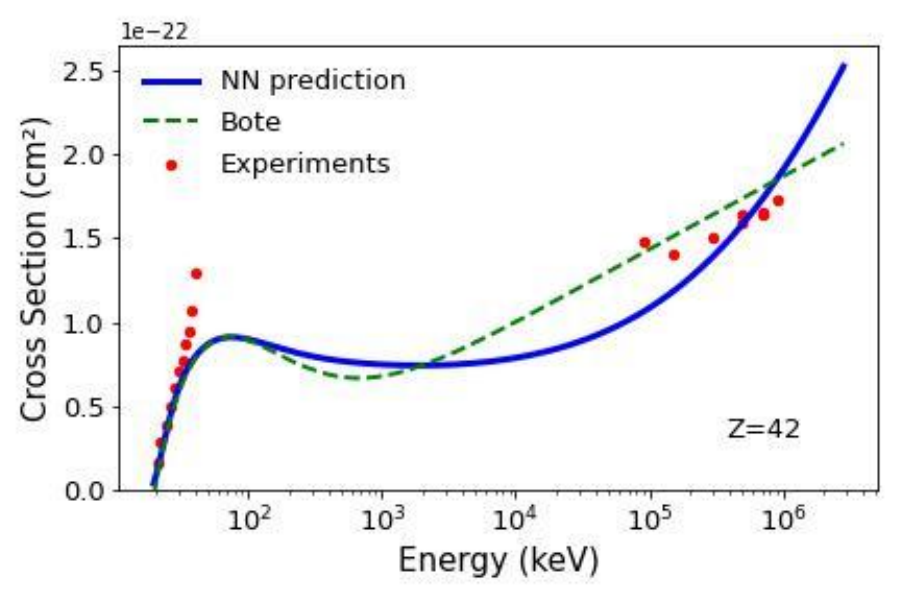}
\hspace{0.02\textwidth}
\includegraphics[width=0.42\textwidth]{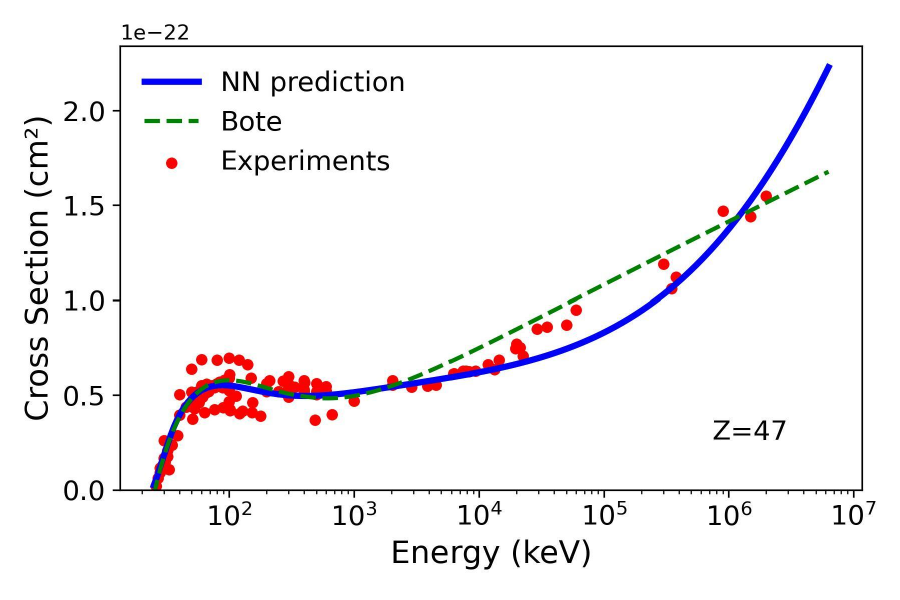} \\
\includegraphics[width=0.42\textwidth]{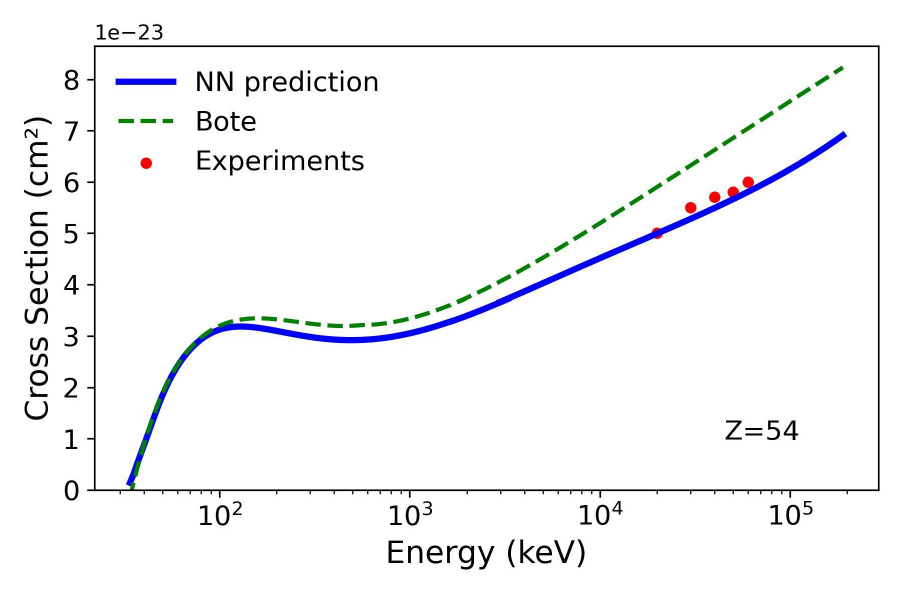}
\hspace{0.02\textwidth}
\includegraphics[width=0.42\textwidth]{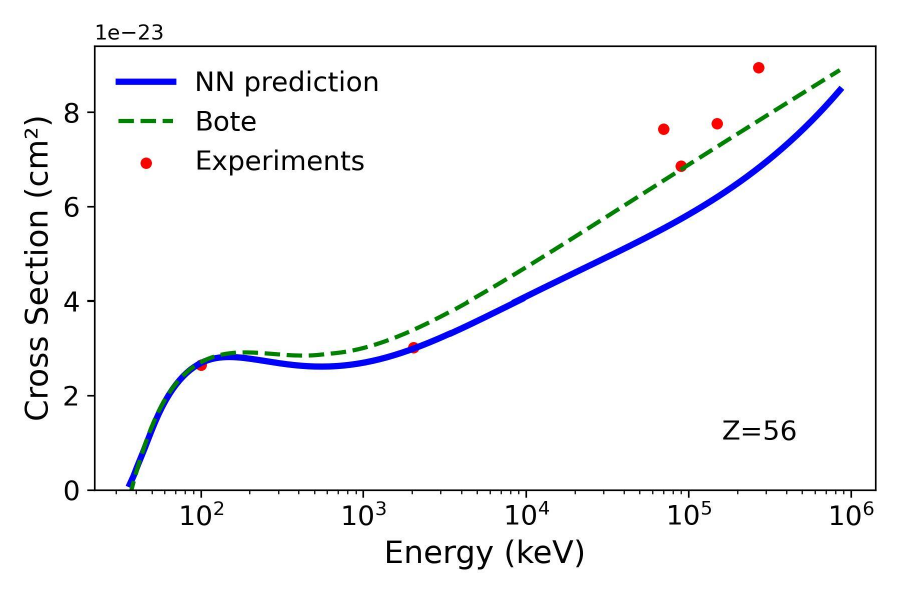} 
\caption{ \raggedright K-shell ionization cross sections as a function of the electron impact-energy for Mo, Ag, Xe, and Ba. 
Curves and symbols as in Figure \ref{fig:KlowZ}.}
\label{fig:KMoAgXeBa}
\end{figure*}

It can be observed that the standard deviation in the values of $\log\sigma_K$, which are the data used to train the neural network, is always within 0.5\%, reflecting narrow distributions around the mean value. On the other hand, for the values of $\sigma_K$, a standard deviation around 20\% is achieved when outliers are not considered, which is consistent with the critical dispersion of the experimental data (see Figs. \ref{fig:KlowZ}--\ref{fig:KMoAgXeBa}). The 1\% of the data lying at each end of the distribution are considered outliers and ignored in the analysis shown in the fourth and sixth columns. As can be seen, excluding a few departed data points results in a drastic decrease in the standard deviation in all cases.

\begin{figure*}[hbt]
\centering
\includegraphics[width=0.32\textwidth]{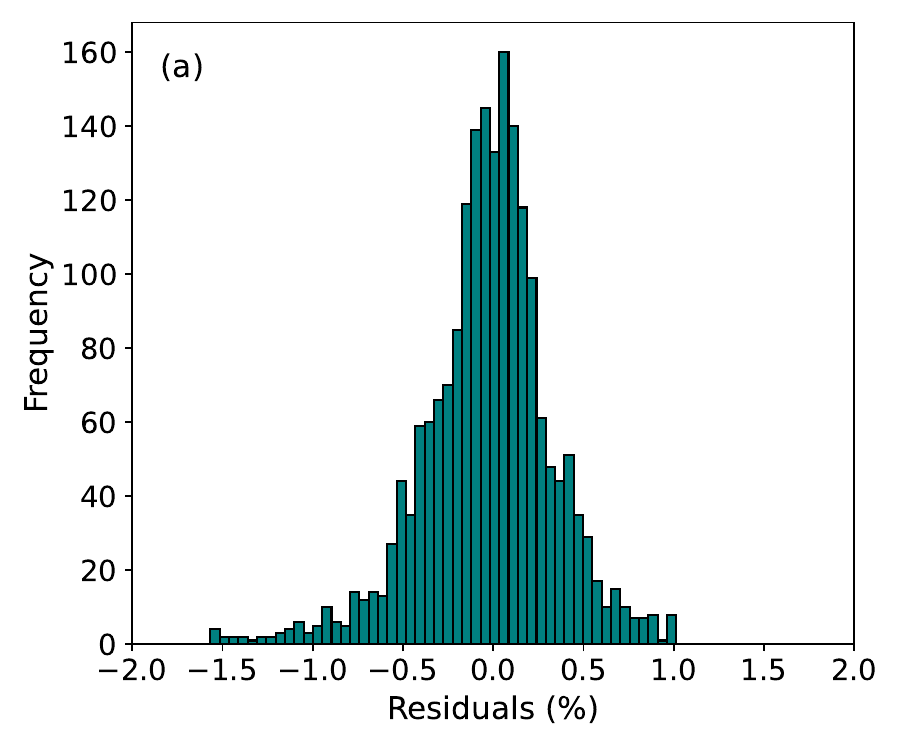}
\includegraphics[width=0.32\textwidth]{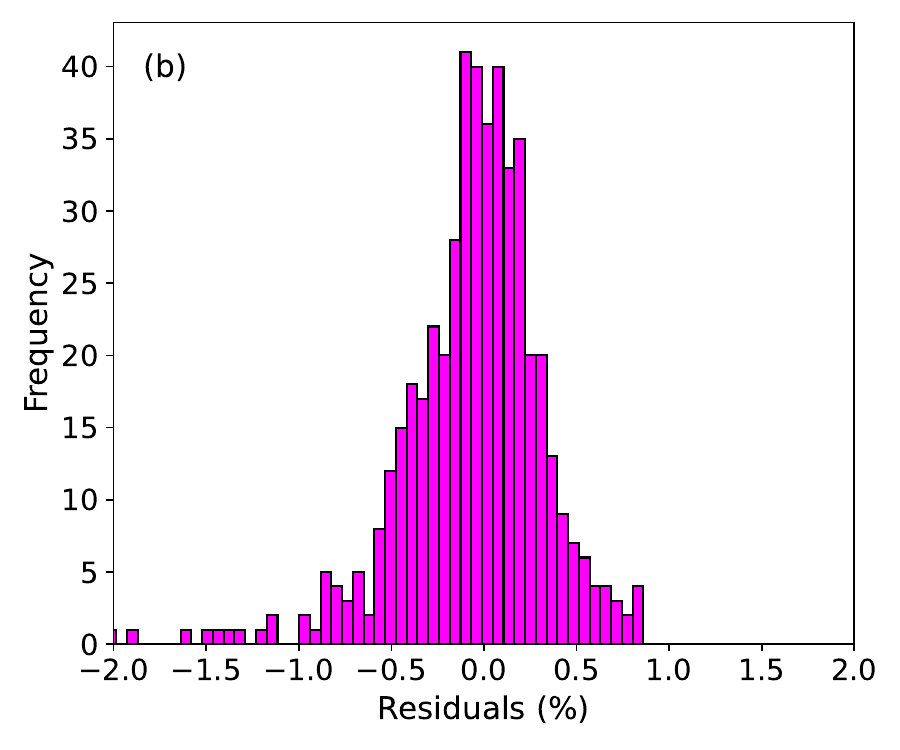}
\includegraphics[width=0.32\textwidth]{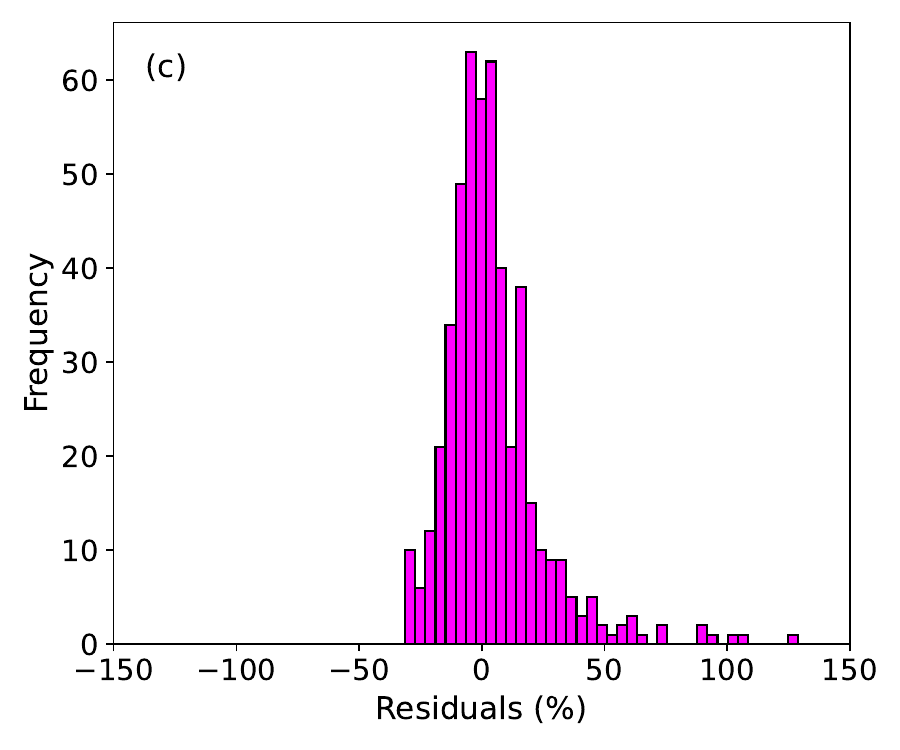}
\caption{\raggedright Histogram of residuals corresponding to the dataset without 1\% of the data from each end of the distribution. 
(a) Residuals of $\log\sigma_K$ in the training dataset; (b) residuals of $\log\sigma_K$ in the test dataset; (c)  residuals of $\sigma_K$ in the test dataset.}
\label{histo} 
\end{figure*}

Figure~\ref{histo} shows residual histograms, where symmetric distributions centered around zero can be observed, which confirms the unbiased behavior of the model. The histograms plotted in Figs.~\ref{histo} (a) and (b) show the distribution of $\log\sigma_K$ residuals in the training and test datasets, respectively. As can be seen, the distribution of residuals obtained with the test dataset is very similar to that achieved with the training dataset, which supports the reliability of the {\sc ikebana} model. The histogram of Fig.~\ref{histo} (c) corresponds to the distribution of residuals for $\sigma_K$ values of the test dataset and shows the quality of the ionization cross-section predictions, which is in line with the dispersion of the available experimental data.

\begin{figure*}[h ]
\includegraphics[width=0.42\textwidth]{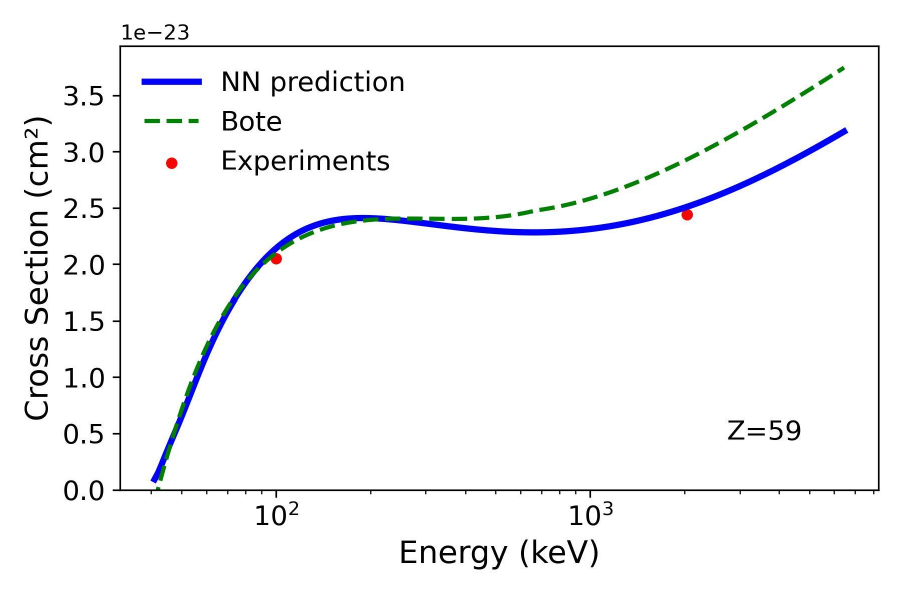}
\hspace{0.02\textwidth}
\includegraphics[width=0.42\textwidth]{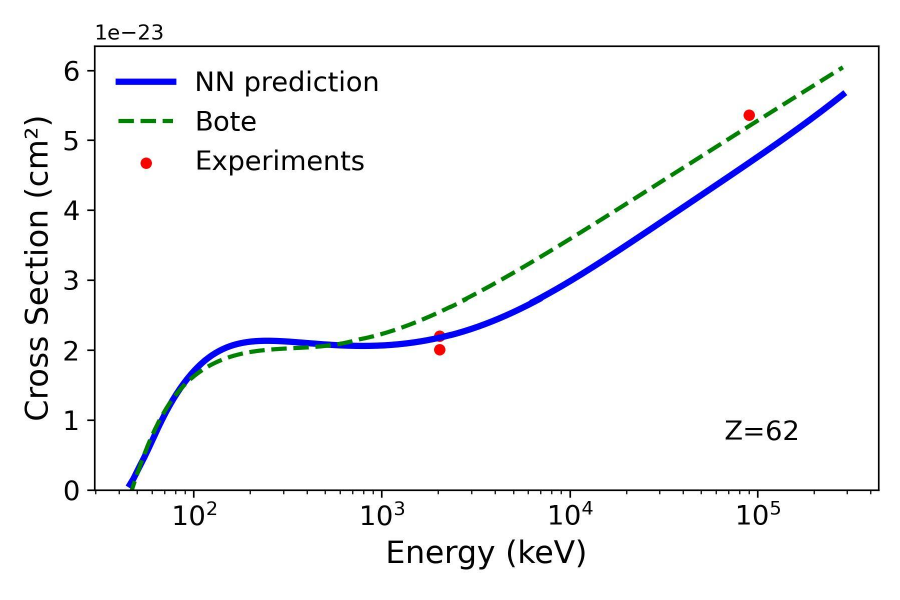} \\
\includegraphics[width=0.42\textwidth]{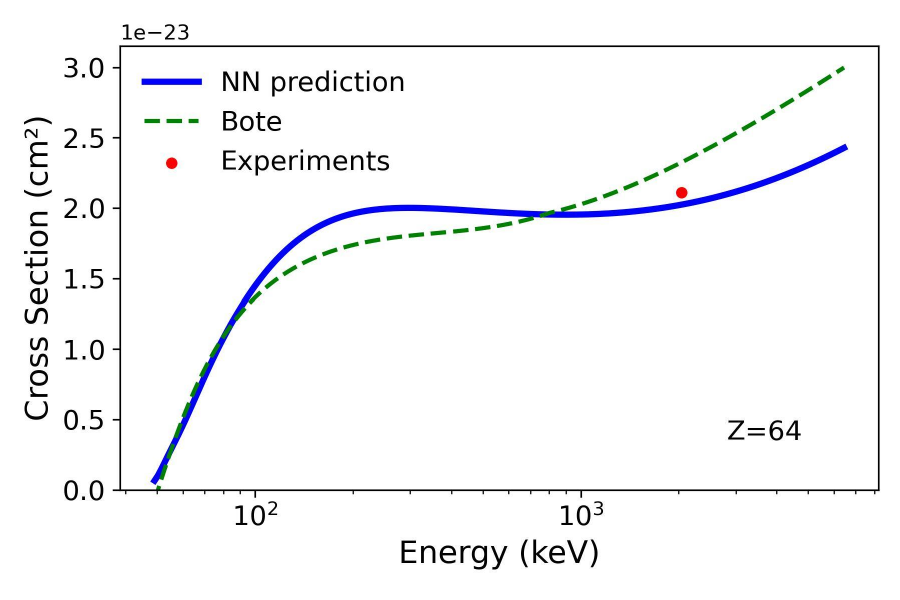}
\hspace{0.02\textwidth}
\includegraphics[width=0.42\textwidth]{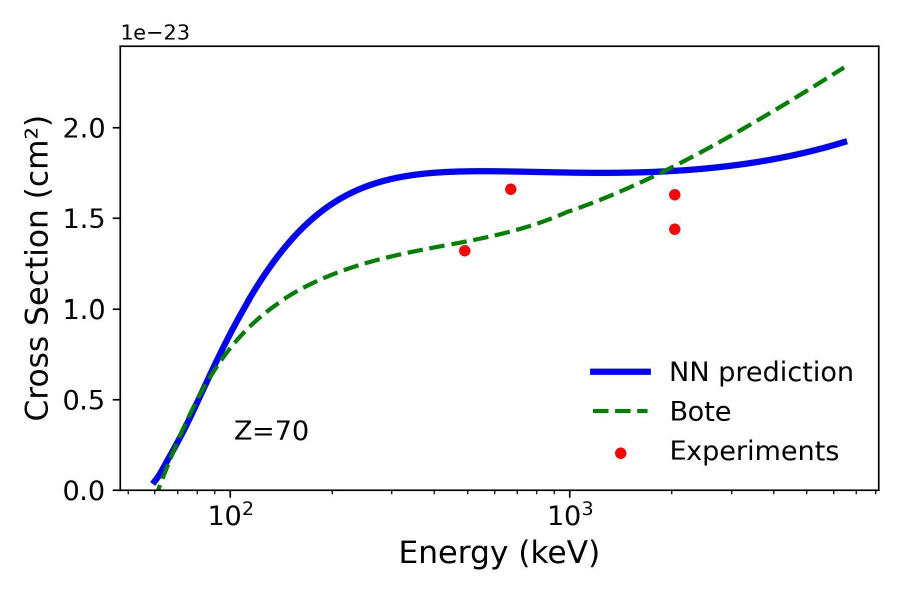} 
\caption{ \raggedright K-shell ionization cross sections as a function of the electron impact-energy for the lanthanides 
Pr, Sm, Gd, and Yb.
Curves and symbols as in Figure \ref{fig:KlowZ}.}
\label{fig:Klanthanides}
\end{figure*}

\begin{figure*}[h ] 
\includegraphics[width=0.42\textwidth]{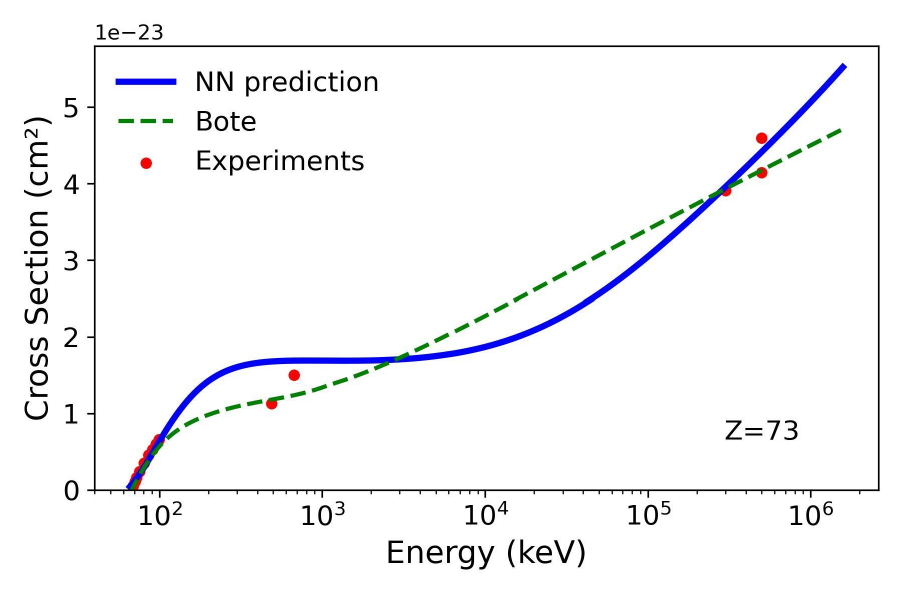}
\hspace{0.02\textwidth}
\includegraphics[width=0.42\textwidth]{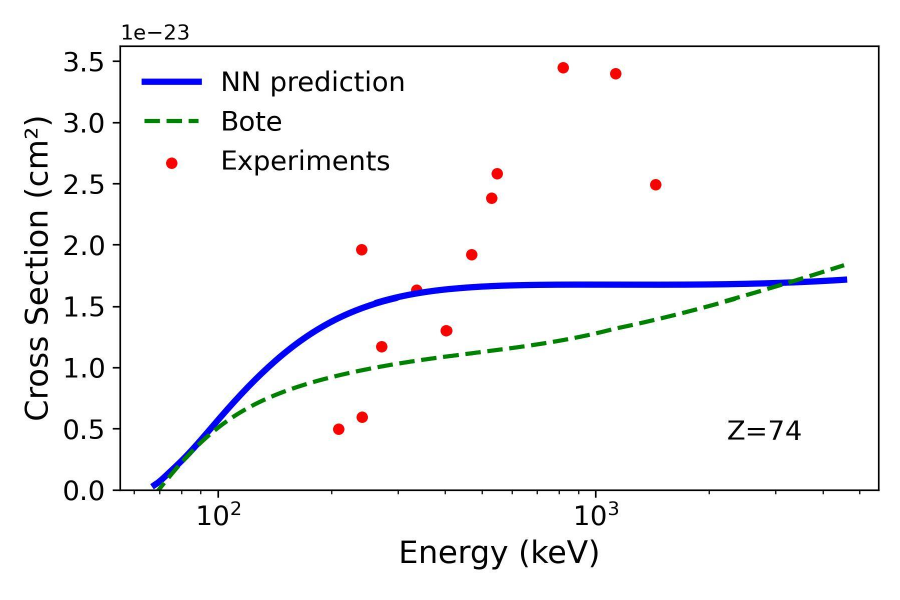} \\
\includegraphics[width=0.42\textwidth]{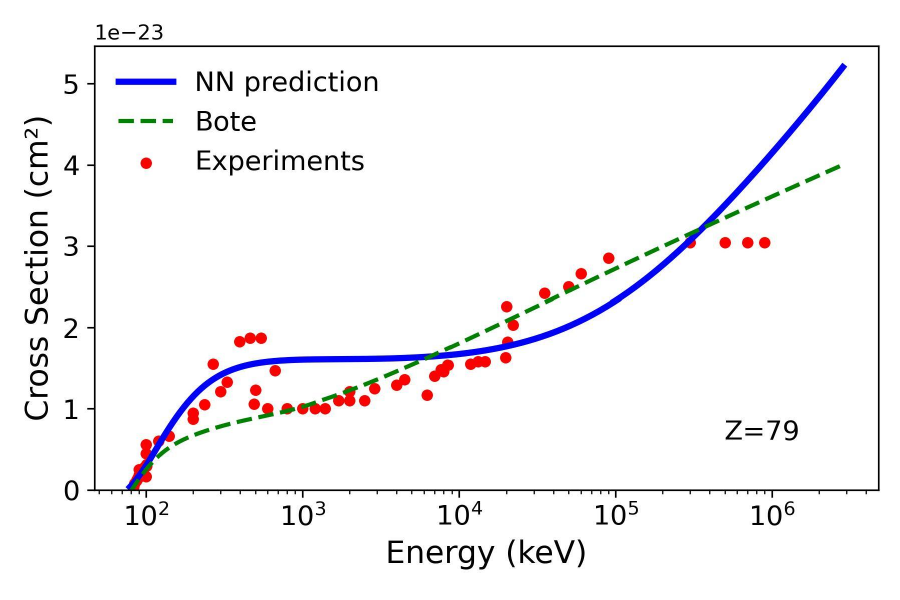}
\hspace{0.02\textwidth}
\includegraphics[width=0.42\textwidth]{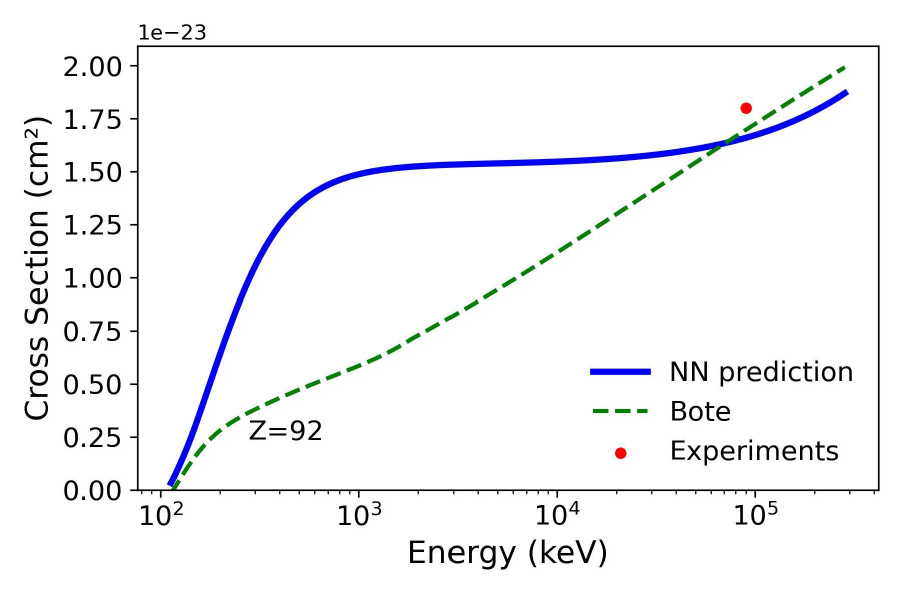} 
\caption{ \raggedright K-shell ionization cross sections as a function of the electron impact-energy for the heavy elements Ta, W, Au, and U. 
Curves and symbols as in Figure \ref{fig:KlowZ}.}
\label{fig:Kheavy}
\end{figure*}

\begin{figure*}[h ]
\includegraphics[width=0.42\textwidth]{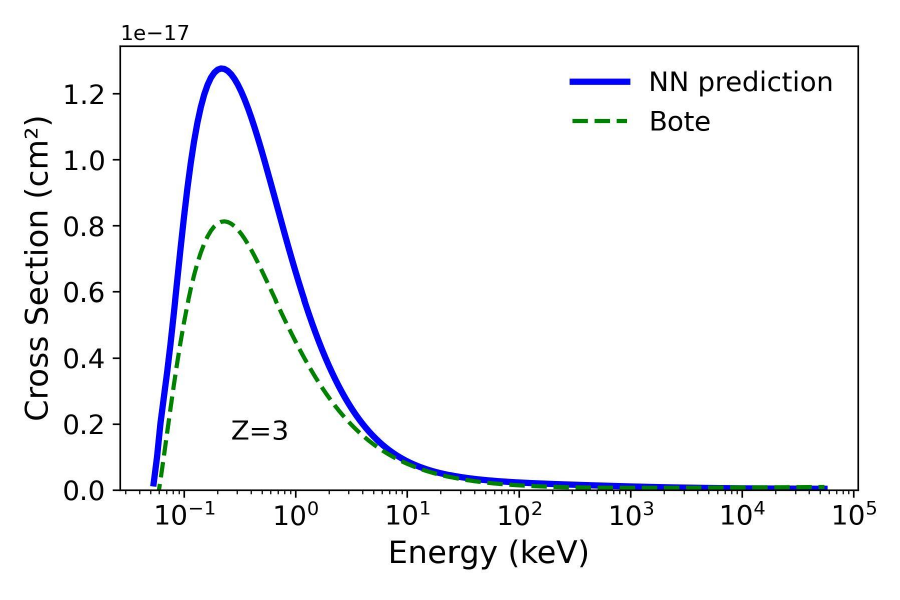}
\hspace{0.02\textwidth}
\includegraphics[width=0.42\textwidth]{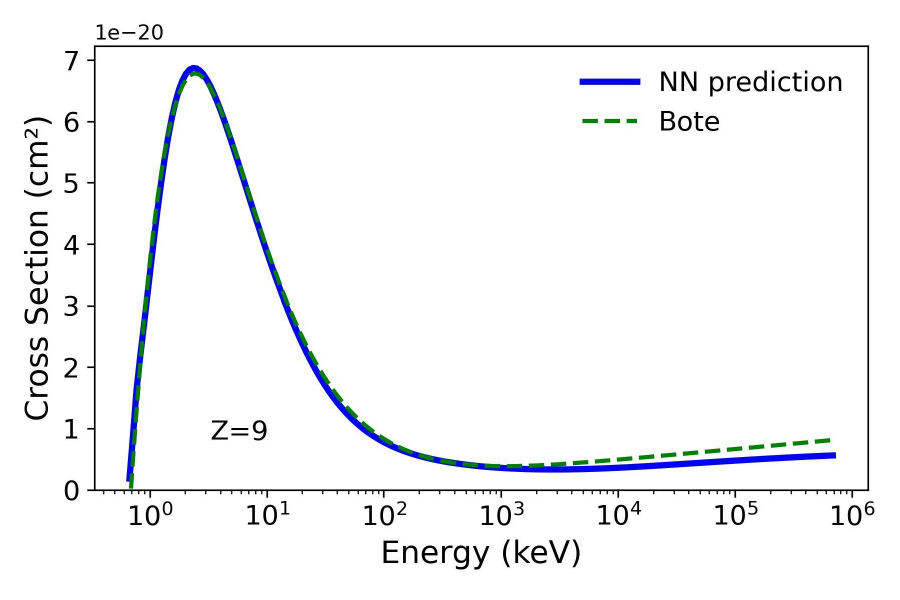} \\
\includegraphics[width=0.42\textwidth]{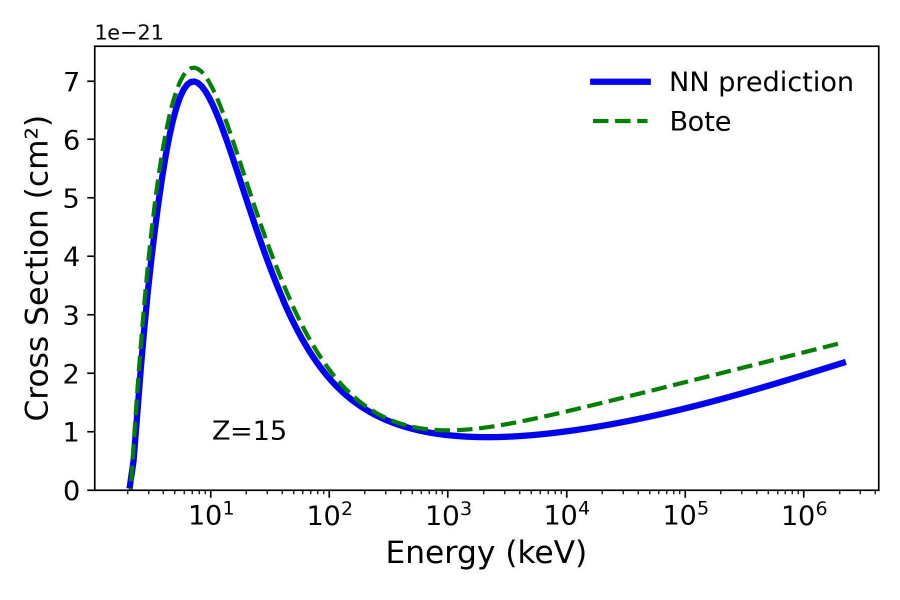}
\hspace{0.02\textwidth}
\includegraphics[width=0.42\textwidth]{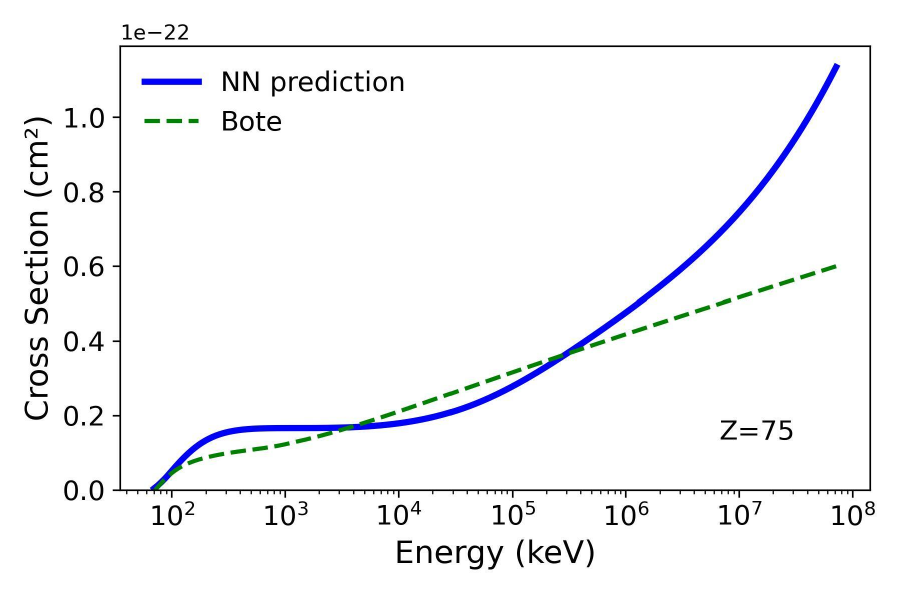} 
\caption{ \raggedright K-shell ionization cross sections as a function of the electron impact-energy for Li, F, P, and Re, with no measurements. 
Curves as in Figure \ref{fig:KlowZ}.}
\label{fig:Knodata}
\end{figure*}

Figures \ref{fig:KlowZ}-\ref{fig:Kheavy} show the $\sigma_K$  predictions as a function of the electron impact energy obtained with the present {\sc ikebana} code for targets ranging from H to U. 
These values are compared with the updated experimental data compilation in \cite{database:24}.
In addition, the theoretical predictions given by Bote {\it et al.} \cite{bote09} are also displayed for comparison, where the distorted-wave Born approximation (DWBA) is used for the lower energies and the plane-wave Born approximation (PWBA) for higher values. It is interesting to note that, although only the experimental data were used to train the network, the results obtained with the neural network show a behavior similar to that of the theoretical approach.
A selection of targets was made to illustrate and discuss the different behaviors observed between model predictions and experimental data.
Additionally, Fig.~\ref{fig:Knodata} presents the current predictions, along with the values reported by Bote, for four representative targets that were not included in the training of the neural network model due to the absence of available experimental data.

Figure~\ref{fig:KlowZ} presents the {\sc ikebana} results for hydrogen, helium, carbon, and oxygen, which are among the most frequently measured targets over time.
The upper-left panel shows the predictions for hydrogen.
In our database, we have identified six studies that report measurements of K-shell ionization cross sections for hydrogen. These results can be grouped into three categories. 
The first group is represented by the measurements by Schram {\it et al.} \cite{Schram:65,Schram:66} and Shyn \cite{Shyn:92}, which overestimate the rest of the experimental cross-section values, particularly at high energies. 
Unlike the other datasets, the experimental values in this group show significant deviations from the theoretical predictions of the PWBA~\cite{bote09}.
The remaining measurements of K-shell ionization of H are consistent at both low and high energies, showing a clear peak in $\sigma_K$ around 60 eV. However, discrepancies remain in the magnitude of this peak. The studies by Shah {\it et al.} \cite{Shah:87} and Rothe {\it et al.} \cite{Rothe:62} report a maximum value of $6 \times 10^{-17}$ cm$^2$, while Fite {\it et al.}~\cite{Fite:58} report a higher peak of $7 \times 10^{-17}$ cm$^2$.
The predictions obtained with our neural network agree with the experimental data across the whole energy range. In particular, they closely reproduce the peak reported by Fite {\it et al.}~\cite{Fite:58}.
This overall agreement suggests that the neural network approach provides a reliable description of K-shell ionization for hydrogen. In contrast, the theoretical calculations by Bote {\it et al.}~\cite{bote09} show significant discrepancies with all experimental values from the ionization threshold up to the peak, which this approximation places at 40 eV. Agreement with the majority of experimental data only begins to emerge above 100 eV.

The helium case, shown in the next panel of Fig.~\ref{fig:KlowZ}, is particularly significant due to the wide availability of experimental results. In contrast to hydrogen, the experimental data for helium show much less dispersion. The predictions from our neural network lie in very good agreement with the average values of the experimental results across the entire energy range. Once again, the theoretical calculations by Bote {\it et al.}~\cite{bote09} match the experimental data at high energies but overestimate the cross sections at low energies, even beyond the maximum.

The lower panels of Fig.~\ref{fig:KlowZ} display the $\sigma_K$ values for carbon (left) and oxygen (right). For these two elements, we observe that our predictions are in good agreement with the experimental data at both low and high energies. However, discrepancies arise in both the position and magnitude of the peak. It is also worth noting that the approximation results provided by Bote {\it et al.}~\cite{bote09} fail to reproduce the experimental data, yielding a curve that resembles our neural network predictions, especially in the case of oxygen, but without matching the experimental values.

Figure~\ref{fig:KAlClArCa} displays the {\sc ikebana} results for aluminum, chlorine, argon, and titanium. These four cases span a wide energy range, from ionization thresholds of a few keV up to the relativistic GeV region.
All curves exhibit a similar general trend, with very good agreement between the predictions of our neural network and the theoretical calculations by Bote {\it et al.}\cite{bote09}.
There is also an overall consistency between the two methods in both the position and magnitude of the cross-section maxima.
In the case of aluminum ($Z=13$), the agreement between our {\sc ikebana} predictions and the experimental data is remarkable across the entire energy range.
This is not the case for chlorine ($Z=17$), where the experimental results reported by Wu {\it et al.}\cite{wu11} are approximately 25\% lower than the theoretical values.
For argon ($Z=18$), our {\sc ikebana} results show very good agreement with the experimental data near the ionization threshold, around the peak, and at high energies. In contrast, the data published by Aydinol~\cite{aydinol07} deviate noticeably from the general trend near the maximum of $\sigma_K$.
In the case of titanium ($Z=22$), both our {\sc ikebana} predictions and the theoretical results by Bote {\it et al.}~\cite{bote09} provide a good description of a specific group of data points near the peak and at low energies. 
However, they disagree with the lower-energy measurements reported by He {\it et al.}~\cite{he97}, which show a systematic deviation from the rest of the dataset.

The results for chromium ($Z=24$), nickel ($Z=28$), copper ($Z=29$), and germanium ($Z=32$), shown in Fig.~\ref{fig:KCrNiCuGe}, reinforce the trends observed in previous cases and confirm the consistency of our predictions across a broad energy range, from threshold up to the GeV region.
A logarithmic scale is used in these plots to better visualize the agreement near the ionization threshold, where fine details tend to be lost in linear representations.
The level of agreement between our {\sc ikebana} predictions and the experimental data is comparable to the intrinsic dispersion observed among the measurements themselves.
The agreement with the theoretical calculations by Bote {\it et al.}~\cite{bote09} is also satisfactory throughout most of the energy range. However, some discrepancies appear in the high-energy asymptotic behavior. These deviations might be attributed to the limited availability of experimental data in that region, which poses a challenge for the predictive power of data-driven models such as {\sc ikebana}.
Overall, we hope that the present study, along with comparisons against independent theoretical approaches, can contribute to a clearer understanding of the existing experimental landscape and emphasize the pressing need for new measurements, particularly at high energies, across a broader range of elements.

The upper-right panel of Fig.~\ref{fig:KMoAgXeBa} shows the K-shell ionization cross sections for silver ($Z=47$), a particularly valuable case due to the abundance of experimental measurements spanning a broad energy range, from the ionization threshold near 20 eV up to approximately 2 GeV. The predictions from our {\sc ikebana} model closely reproduce the experimental data across this entire range, except for a few isolated points around 40 MeV. Although the experimental trend follows the asymptotic behavior predicted by the theoretical calculations of Bote {\it et al.}\cite{bote09}, it also aligns well with our predictions over a broad intermediate energy region where {\sc ikebana} and Bote slightly diverge.
Figure~\ref{fig:KMoAgXeBa} also displays the predicted $\sigma_K$ values for molybdenum ($Z=42$), xenon ($Z=54$), and barium ($Z=56$), elements for which only limited experimental data are available. In the case of molybdenum, measurements exist both near the ionization threshold and in the high-energy range between $10^5$ and $10^6$ keV. At low energies, our predictions show excellent agreement with both experimental data and the theoretical values from Bote {\it et al.} However, the predicted maxima differ significantly from the experimental trend, and the precise location of the peak remains uncertain. Notably, the {\sc ikebana} and Bote curves coincide in both the position and value of the peak, despite the fundamental differences between the two approaches.
For xenon, only five experimental data points are available, all of which are in the relativistic energy regime. These points match remarkably well with our neural network predictions. In contrast, they deviate considerably from the results of Bote {\it et al.} Finally, in the case of barium, our {\sc ikebana} model reproduces the two available data points at 0.1 and 2 MeV, but underestimates the values reported by Ishii {\it et al.}\cite{ishii77} around 100~MeV.

The case of rare earth elements, also known as lanthanides, has gained increasing attention in recent years due to their diverse range of technological applications~\cite{Patil:22}. However, experimental data for K-shell ionization cross sections remain extremely scarce. Figure~\ref{fig:Klanthanides} displays four representative examples: praseodymium ($Z=59$), samarium ($Z=62$), gadolinium ($Z=64$), and ytterbium ($Z=70$). Despite the limited number of available data points, our {\sc ikebana} predictions show good agreement with the measurements, offering a valuable alternative for estimating missing values in this poorly explored region of the periodic table.

Figure~\ref{fig:Kheavy} presents the K-shell ionization cross sections for some of the heaviest targets in our study: tantalum ($Z=73$), tungsten ($Z=74$), gold ($Z=79$), and uranium ($Z=92$). The {\sc ikebana} predictions for tantalum show excellent agreement with the experimental data near the ionization threshold and in the high-energy region.
Tungsten has attracted considerable attention in recent decades, particularly due to its selection as the first-wall material for the International Thermonuclear Experimental Reactor (ITER)\cite{ITER}. However, this interest has not extended to K-shell ionization studies, and the available data are both limited in energy coverage and highly scattered. All measurements for W stem from the work of Hansen {\it et al.}\cite{hansen64,hansen66} in the 1960s.
Gold, as expected, has received more sustained experimental focus. Measurements span a wide energy range, and our {\sc ikebana} predictions show strong agreement with these data, remaining within their reported experimental uncertainties. However, the predictions of Bote {\it et al.}~\cite{bote09} appear to provide a slightly better description of the experimental trend across the full range.
Finally, for uranium, only a single experimental point is available, reported by Ishii {\it et al.} in 1977~\cite{ishii77}. Our {\sc ikebana} model accurately reproduces this value. Nevertheless, we emphasize the importance of the behavior at lower energies, predicted by both our model and that of Bote {\it et al.}, in a regime where no experimental data currently exist.

As noted throughout this discussion, experimental data remain unavailable for a significant number of targets. Specifically, there are 27 elements between hydrogen and uranium for which no K-shell ionization cross-section measurements exist, and 13 additional elements with at most two reported data points.
Despite this scarcity, our {\sc ikebana} model provides continuous predictions for all elements across the periodic table.
Figure~\ref{fig:Knodata} illustrates examples for four such unmeasured elements: lithium ($Z=3$), fluorine ($Z=9$), phosphorus ($Z=15$), and rhenium ($Z=75$). The ability of the neural network to interpolate and generalize from the training data corresponding to adjacent elements enhances the credibility of these predictions.
Experimental measurements for these targets would be highly valuable, both to validate existing predictions and to further refine the accuracy of data-driven models such as {\sc ikebana}.

\section{Conclusions}
\label{sec:conclusions}

This work presents a machine learning model for predicting electron-impact K-shell ionization cross sections, trained on a recent, comprehensive, and updated compilation of experimental data~\cite{database:24}.
The resulting neural network, {\sc ikebana}, exhibits excellent predictive performance, as demonstrated by its agreement with the experimental data in both training and test datasets.
It provides accurate cross-section values across a wide energy range, from the ionization threshold to the peak and well into the relativistic regime, and shows strong consistency with available experimental data.
Importantly, {\sc ikebana} delivers reliable predictions in energy regions where no measurements exist, and for a significant number of atomic targets for which no experimental data are currently available. In many cases, the model results show better agreement with experimental data than existing theoretical approaches, such as those by 
Bote et al.~\cite{bote09}, particularly near threshold energies and around the cross-section maxima.
Overall, the model offers a powerful and flexible tool for estimating ionization cross sections across the periodic table, complementing theoretical calculations and guiding future experimental efforts. Additional measurements, especially for intermediate and high-Z elements at low energies, would be valuable for validating and further refining the predictive capabilities of the model.

\begin{acknowledgments}
The following institutions of Argentina financially support this research:
the Consejo Nacional de Investigaciones cient\'{\i}ficas y T\'ecnicas (CONICET) by the projects PIP11220200102421CO and  PIP11220200100986CO, the Agencia Nacional de Promoci\'on Cient\'{\i}fica y Tecnol\'ogica (ANPCyT) by the project PICT-2020-SERIE A-01931, and the Secretar\'{\i}a de Ciencia y T\'ecnica de la Universidad
Nacional de C\'ordoba, by the project 33620230100222CB.
\end{acknowledgments}
~  

\section*{Data availability}
The {\sc ikebana} code developed can be freely used in the Google Colab platform:   \href{https://colab.research.google.com/drive/1_ko9GtoOEifZ44mZYICoTcvFEuWBIdIm?usp=copy}{{\it K-ionization with Machine Learning}}\cite{ikebana:25}. 
The user has to execute the cells and only needs to specify the atomic number $Z$ and the desired range of incident electron energies, defining $E_{min}$ and $E_{max}$. The notebook will generate plots comparing the results with the available data and tables containing the corresponding K-shell ionization cross-section results. Tables and plots can be downloaded from the folder "Results".
In addition, the data that support the findings of this study are openly available in \href{https://www.famaf.unc.edu.ar/~trincavelli/database.html}{{\it Cross sections for K-shell ionization by electron impact database}}\cite{database:24}.

\section{References}
\label{sec:references}



\end{document}